\documentclass[onecolumn]{aa} % for a paper on 1 column
\usepackage{graphicx}
\usepackage{txfonts}
\begin{document}

   \title{Preprocessing of vector magnetograms for magnetohydrostatic extrapolations}

   \author{X. Zhu\inst{1}, T. Wiegelmann\inst{1}, and B. Inhester\inst{1}}

  \institute{Max-Planck-Institut f\"{u}r Sonnensystemforschung, Justus-von-Liebig-Weg 3, 37077 G\"{o}ttingen, Germany\\
            \email{zhu@mps.mpg.de}}

   \date{Received ; accepted }
%   \date{Received September 15, 1996; accepted March 16, 1997}

% \abstract{}{}{}{}{}
% 5 {} token are mandatory
% , whereas some model assumptions are necessary
  \abstract
   {Understanding the 3D magnetic field as well as the plasma in the chromosphere and transition region is important. One way is to extrapolate the magnetic field and plasma from the routinely measured vector magnetogram on the photosphere based on the
assumption of the magnetohydrostatic (MHS) state. However, photospheric data may be inconsistent with the MHS assumption. Therefore, we must study the restriction on the photospheric magnetic field, which is required by the MHS system. Moreover, the data should be transformed accordingly before MHS extrapolations can be applied.}
   {We aim to obtain a set of surface integrals as criteria for the MHS system and use this set of integrals to preprocess a vector magnetogram.}
   {By applying Gauss' theorem and assuming an isolated active region on the Sun, we related the magnetic energy and forces in the volume to the surface integral on the photosphere. The same method was applied to obtain restrictions on the photospheric magnetic field as necessary criteria for a MHS system. We used an optimization method to preprocess the data to minimize the deviation from the criteria as well as the measured value.}
   {By applying the virial theorem to the active region, we find the boundary integral that is used to compute the energy of a force-free field usually underestimates the magnetic energy of a large active region. We also find that the MHS assumption only requires the x-, y-component of net Lorentz force and the z-component of net torque to be zero. These zero components are part of Aly's criteria for a force-free field. However, other components of net force and torque can be non-zero values. According to new criteria, we preprocess the magnetogram to make it more consistent with the MHS system and, at the same time close, to the original data.}
   {}

\titlerunning{Magnetogram preprocessing for a MHS system}
\maketitle

%-------------------------------------------------------------------
\section{Introduction}

%\begin{align}
%\begin{split}
%a&={}b+c+d+e+(f+g)+h\\
%&+d+e,
%\end{split}\\
%\begin{split}
%a+b+c+d&={}c+d+e\\
%&f
%\end{split}\\
%a+d&={}c
%\end{align}

The force-free assumption $\nabla \times \bf{B}=\alpha(\bf r)B$ has long been the basis for a magnetic field extrapolation from the solar photosphere to the corona \citep{ws12}. As a result, a number of force-free field extrapolation techniques have been developed over half a century. The simplest case is the potential field or current-free field extrapolation in which $\alpha=0$ \citep{s64,swn69}. A next step with a spatially constant $\alpha$ is the linear force-free field extrapolation \citep{ch77,s78,a81}. At last, a model with non-constant $\alpha$ is called the nonlinear force-free field (NLFFF) extrapolation. The NLFFF extrapolations include the (1) upward integration method \citep{n74,wch85,wsc90,cds91,dp92,sft06}, (2) Grad-Rubin method \citep{gr58,s81,aal97,abm99,aba06,w04a,w06}, (3) relaxation method \citep{mbs88,r96,v05,jf12,fwh11,gxk16}, (4) optimization method \citep{wsr00,w04b,wi10}, (5) boundary-element method \citep{ys00,yl06,hw06} and (6) forward-fitting method \citep{msd12,a13,a16}.

In a force-free model, some necessary conditions of the magnetic field on the boundary have to be fulfilled. \cite{m69}, \cite{m74}, and \cite{a84,a89} obtained several integral relations on the boundary corresponding to the vanishing net magnetic force and vanishing torque, respectively. If these integral relations are not fulfilled then the boundary is not consistent with a force-free field. Based on this, \cite{wis06} proposed a preprocessing algorithm to modify the magnetogram within the error margins of the measurements to minimize the net magnetic force and torque. The resulting vector magnetogram is more suitable for a force-free extrapolation. Further developments include using the method of simulated annealing \citep{fsv07}, extending to the spherical geometry \citep{twi09}, adding a new term concerning chromospheric longitudinal fields \citep{yk12}, and dealing with the potential and non-potential components separately \citep{jf14}.

With increasing spatial resolution of the measured vector magnetogram, we can study the magnetic field in the lower solar atmosphere in detail. However, the force-free assumption is not valid anymore in this layer as the plasma $\beta$ is much larger than it is in the corona \citep{g01}. A straightforward approach is to take into account the plasma in the extrapolation. \cite{zwd13}, \cite{zwd16}, and \cite{mki20} proposed the magnetohydrodynamic relaxation method to obtain a magnetohydrostatic (MHS) solution with a non-force-free layer near the bottom boundary. \cite{gw13} and \cite{gbb16} extended the \cite{gr58} method to compute the MHS equilibria. \cite{wn06} used the optimization method to solve the MHS equations without gravitational force in the Cartesian coordinate system. \cite{wnr07} extended the code with gravitational force in spherical coordinate system. Recently, we extended the optimization method by introducing the gravitational force \citep{zw18} in the Cartesian coordinate system, tested the code with an realistic radiative MHD simulation \citep{zw19}, and also applied the code to a \textsc{Sunrise}/IMaX dataset \citep{zws20}. It is worth noting that the new algorithm ensures the positive definiteness of gas pressure and mass density.
%\cite{zw19} suggested \citep[e.g.,][]{srb17}  (e.g., 120km for SST/SOUP, 100 km for SUNRISE/IMaX, and )

As a result of the MHS assumption, we can also define several integral relations just as we did for the force-free field. A preprocessing algorithm (extended from the NLFFF case) is proposed to modify the vector magnetogram within the error margins of the measurement. The resulting magnetogram is expected to be more consistent with the assumption of a MHS extrapolation.

The remainder of the paper is organized as follows: In Sect.~\ref{sec:integrals} we define the integral relations for a MHS system and also apply the virial theorem to an active region. In Sect.~\ref{sec:preprocess} we describe the optimization algorithm to derive consistent boundary conditions for a MHS extrapolation. In Sect.~\ref{sec:application} we apply the algorithm to an example of the observed vector magnetogram. In Sect.~\ref{sec:conclusion} we draw conclusions.

\section{Boundary integrals of a MHS system}\label{sec:integrals}

A MHS equilibrium can be described as
\begin{eqnarray}
(\nabla \times \mathbf{B})\times \mathbf{B}-\nabla p - \rho \mathbf{\hat{z}} & = & 0, \label{eq:force_balance}\\
\nabla \cdot \mathbf{B} & = & 0,
\end{eqnarray}
where $\mathbf{B}$, $p$, and $\rho$ are the magnetic field, plasma pressure ,and plasma density, respectively. We note that $\bf B$, $p$, and $\rho$ have been appropriately normalized to simplify the equation. For example, for a study that includes photosphere, the following normalization constants are convenient: $\rho_0=2.7\times10^{-1}g/cm^3$ (density), $T_0=6\times 10^3K$ (temperature), $g=2.7\times10^4 cm/s^2$ (gravitational acceleration), $L_0=\frac{\mathcal{R}T_0}{\mu g}=1.8\times10^7 cm$ (length), $p_0=\sqrt{\frac{\rho_0\mathcal{R}T_0}{\mu}}=1.3\times10^5 dyn/cm^2$ (plasma pressure), and $B_0=\sqrt{4\pi p_0}=1.3\times10^3 G$ (magnetic field), where $\mathcal{R}$ is the ideal gas constant.

As discussed in many papers \citep[e.g.,][]{c61,m69,m74,a84}, the above force balance equation may be written in the form analogous to equations of elasticity as follows:
\begin{eqnarray}
\bf \nabla \cdot T = f\label{eq:mhs},
%(\nabla \times \mathbf{B})\times \mathbf{B}-\nabla p - \rho \mathbf{\hat{z}} = \bf F,
\end{eqnarray}
where $T_{ij}=\frac{1}{2}B^2\delta_{ij}-B_iB_j$ is the Maxwellian tensor and $\bf f$ includes forces such as pressure gradient $(-\nabla p$) and gravitational force ($-\rho \bf\hat{z}$).

\subsection{Equations of momenta}

%From Eq.~(\ref{eq:mhs}) we may obtain several other integral relations as restrictions on the field value on the boundary.
Assuming a Cartesian coordinate, integrating Eq.~(\ref{eq:mhs}) over the volume V with surface $\bf S$, we get
\begin{eqnarray}
\oint_{\bf S} \left[\left(\frac{1}{2}B^2+p\right){\bf\hat{x}}-B_x{\bf B}\right]\cdot d{\bf s} &=& 0\label{eq:netforceAx}, \\
\oint_{\bf S} \left[\left(\frac{1}{2}B^2+p\right){\bf\hat{y}}-B_y{\bf B}\right]\cdot d{\bf s} &=& 0\label{eq:netforceAy}, \\
\oint_{\bf S} \left[\left(\frac{1}{2}B^2+p\right){\bf\hat{z}}-B_z{\bf B}\right]\cdot d{\bf s}+\int_V \rho dv &=& 0\label{eq:netforceAz}.
\end{eqnarray}
We apply these three relations to an isolated active region with the photosphere as the bottom boundary. As to a closed current system, the magnetic field falls off as $1/r^3$. Therefore as ${\bf r}\rightarrow \infty$, the lateral and top surface integrals with magnetic terms approach zero. We note that in $\int -\nabla pdv=-\oint_{\bf S} pd\bf s$ the transverse surfaces make zero net contribution since, without magnetic fields, the forces acting on the opposite side boundaries by the external plasma pressure are equal and opposite. Then we have net Lorentz force restrictions written as
\begin{eqnarray}
\int_{S_1} B_xB_zdxdy &=& 0\label{eq:netforcex}, \\
\int_{S_1} B_yB_zdxdy &=& 0\label{eq:netforcey}, \\
\int_{S_1} \left(\frac{{B_z}^2-{B_x}^2-{B_y}^2}{2}\right)dxdy-\int_{S_1}pdxdy+\int_V\rho dv &=& 0\label{eq:netforcez},
\end{eqnarray}
where $S_1$ is the bottom boundary. We note that Eqs.~(\ref{eq:netforcex}) and (\ref{eq:netforcey}) are exactly the same as those for a force-free field. However, that is not the case vertically. Eq.~(\ref{eq:netforcez}) implies that the difference between the pressure at the bottom boundary and the weight of the plasma above is compensated by the Lorentz force.

%multiplying Eq.~(\ref{eq:mhs}) vectorially by $\bf r$:

The net Lorentz force torque restrictions can also be obtained in a similar way with the cross product of Eq.~(\ref{eq:mhs}) and $\bf r$ as follows:
\begin{eqnarray}
\int_{S_1} y\left(\frac{{B_x}^2+{B_y}^2-{B_z}^2}{2}+p\right)dxdy + \int_V\rho ydv &=& 0\label{eq:nettorquex}, \\
\int_{S_1} x\left(\frac{{B_x}^2+{B_y}^2-{B_z}^2}{2}+p\right)dxdy + \int_V\rho xdv &=& 0\label{eq:nettorquey}, \\
\int_{S_1} (yB_zB_x-xB_zB_y)dxdy &=& 0\label{eq:nettorquez}.
\end{eqnarray}
We see Eq.~(\ref{eq:nettorquez}) is exactly the same as that for a force-free field, while that is not the case in two transverse directions. Eqs.~(\ref{eq:nettorquex}) and (\ref{eq:nettorquey}) imply that the plasma may induce rotational moments relative to the x or y axes.

\subsection{Virial theorem}

Multiplying Eq.~(\ref{eq:mhs}) by ${\bf r}=x{\bf\hat{x}}+y{\bf\hat{y}}+z{\bf\hat{z}}$ and integrate over volume V with a surface $\bf S$, we derive the virial theorem
\begin{eqnarray}
\int_V\left(\frac{1}{2}B^2+{\bf f\cdot r}\right)dv = \oint_S\left[\frac{1}{2}B^2{\bf r}-{\bf (B\cdot r)B}\right]\cdot d\bf s\label{eq:virial1},\end{eqnarray}
which relates the magnetic energy and forces inside the volume to the integral on the surface. If the electric currents are closed, the magnetic field falls off as $1/r^3$, and as ${\bf r}\rightarrow \infty$ the surface integral approaches zero. Therefore the magnetic energy vanishes if ${\bf f}=0$. That means a force-free field does not exist in an isolated current system \citep{c61}.

Plugging ${\bf f}=-\nabla p-\rho {\bf \hat{z}}$ into Eq.~(\ref{eq:virial1}), we have
\begin{eqnarray}
\int_V\left(E_t+\frac{3\gamma-4}{\gamma-1}p\right)dv = \oint_{\bf S}\left[\left(\frac{1}{2}B^2+p\right){\bf r}-{\bf (B\cdot r)B}\right]\cdot d\bf s\label{eq:virial2},
\end{eqnarray}
where $E_t=\frac{1}{2}B^2+\frac{p}{\gamma-1}-\rho z$ is the total energy including magnetic energy $\frac{1}{2}B^2$, internal energy $\frac{p}{\gamma-1}$ and gravitational potential energy $-\rho z$. If $\gamma=\frac{4}{3}$ then the left-hand side (LHS) of Eq.~(\ref{eq:virial2}) becomes exactly the total energy. Eq.~(\ref{eq:virial2}) relates the total energy and pressure in the volume to the integral on the surface.

\subsection{Apply to an active region}

Let us focus on an active region. Assume a semi-infinite half-space with the bottom boundary on the photopshere, the magnetic field of an isolated active region decreases with distance as $1/r^3$, thus the surface integral in Eq.~(\ref{eq:virial1}) on the side and top boundaries will approach zero. Then we get:
\begin{eqnarray}
\int_V\left(\frac{1}{2}B^2+{\bf f\cdot r}\right) dv = \int_{S_1}\left[{\bf (B\cdot r)}B_z-\frac{1}{2}B^2z\right]dxdy\label{eq:virial4},
\end{eqnarray}
where $S_1$ is the bottom boundary.

\subsubsection{Effect of translation on the integration}\label{sec:trans}

In the force-free case, the integral at the right-hand side (RHS) of Eq.~(\ref{eq:virial4}) is invariant under translation. However, it seems from the LHS of Eq.~(\ref{eq:virial4}) that this nice property is not fulfilled any more in a MHS system. Suppose a translation: $(x,y,z)\longrightarrow(x'+\Delta x,y'+\Delta y,z'+\Delta z)$. Then $\bf B'(r')=B(r)$, $dx'=dx$, and $dy'=dy$. Thus we have the RHS of Eq.~(\ref{eq:virial4}) in the new coordinate system as follows:
\begin{eqnarray}
&&\int_{S_1'}\left[{\bf (B'\cdot r')}{B'}_z-\frac{1}{2}{B'}^2z'\right]dx'dy'\\
=&&\int_{S_1}\left[{\bf B\cdot (r+\Delta r)}{B}_z-\frac{1}{2}{B}^2(z+\Delta z)\right]dxdy\label{eq:trans2}\\
=&&\int_{S_1}\left[{\bf (B\cdot r)}{B}_z-\frac{1}{2}{B}^2z\right]dxdy+\int_{S_1}\left(\frac{{B_z}^2-{B_x}^2-{B_y}^2}{2}\right)\Delta zdxdy\label{eq:trans3},
\end{eqnarray}
where ${\bf \Delta r}=(\Delta x, \Delta y, \Delta z)$. We note that Eqs.~(\ref{eq:netforcex}) and (\ref{eq:netforcey}) are used to infer formula (\ref{eq:trans3}) from (\ref{eq:trans2}). Thus the integration on the bottom boundary is invariant under the translation in the same horizontal plane with $\Delta z=0$.

\subsubsection{Estimate the magnetic energy}

Suppose ${\bf r}=0$ on the bottom boundary, Eq.~(\ref{eq:virial4}) becomes
\begin{eqnarray}
\int_V\left(\frac{1}{2}B^2+{\bf f\cdot r}\right) dv = \int_{S_1}B_z(xB_x+yB_y)dxdy\label{eq:virial3}.
\end{eqnarray}
As with a force-free field, the integral at the RHS of Eq.~(\ref{eq:virial3}) is often used to estimate the magnetic energy in the semi-infinite volume. However, the integral is not equal to the magnetic energy in a MHS equilibrium. In the following analysis, we try to judge, in a MHS equilibrium, whether the integral is an underestimate or an overestimate of the magnetic energy.

\begin{figure}
\centering
\includegraphics[width=\hsize]{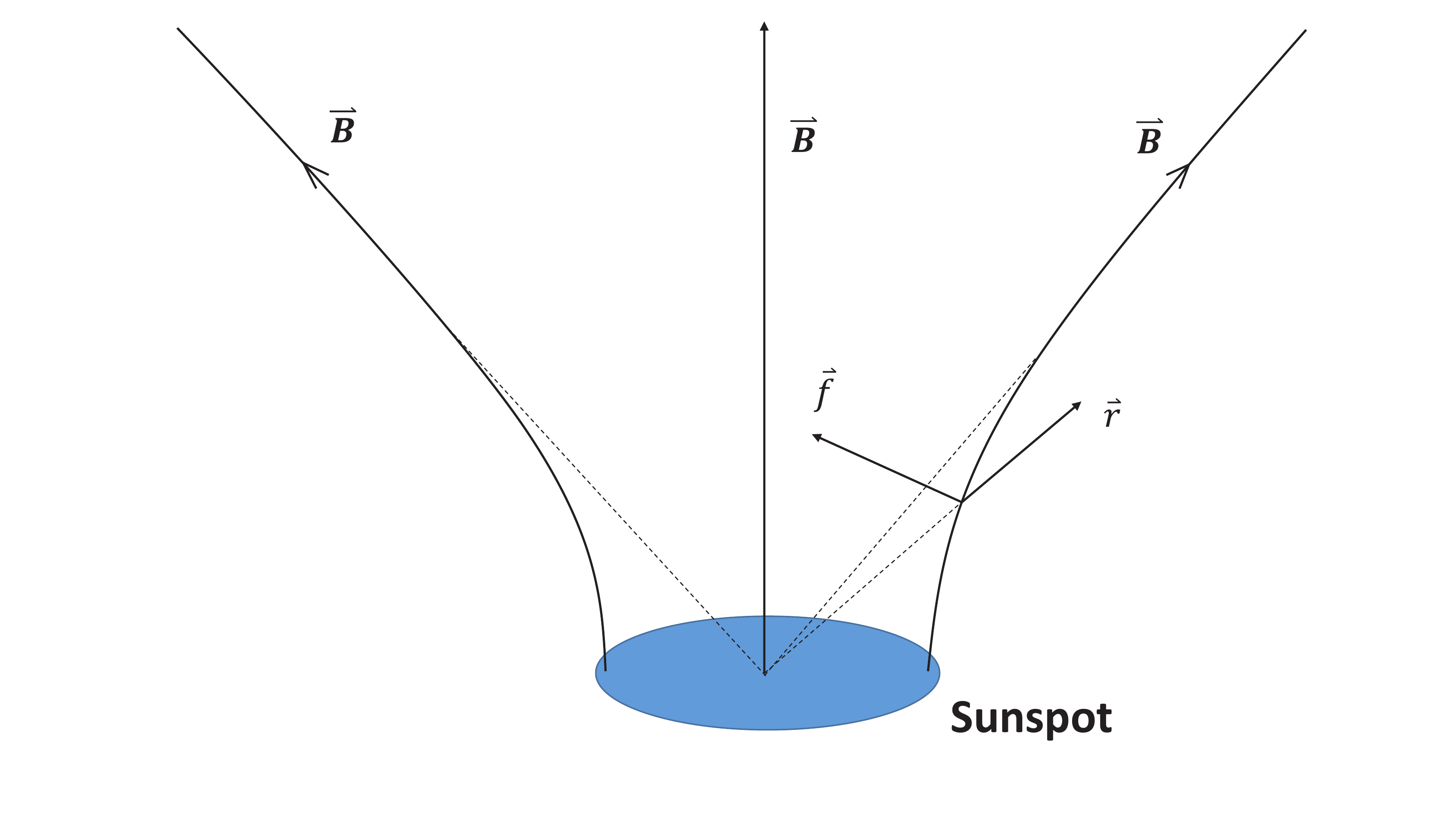}
\caption{Cartoon shows, in a MHS state, a typical interaction of the magnetic force and plasma forces in an active region with single compact polarity.}
\label{fig:fr}
\end{figure}

\begin{figure}
\centering
\includegraphics[width=\hsize]{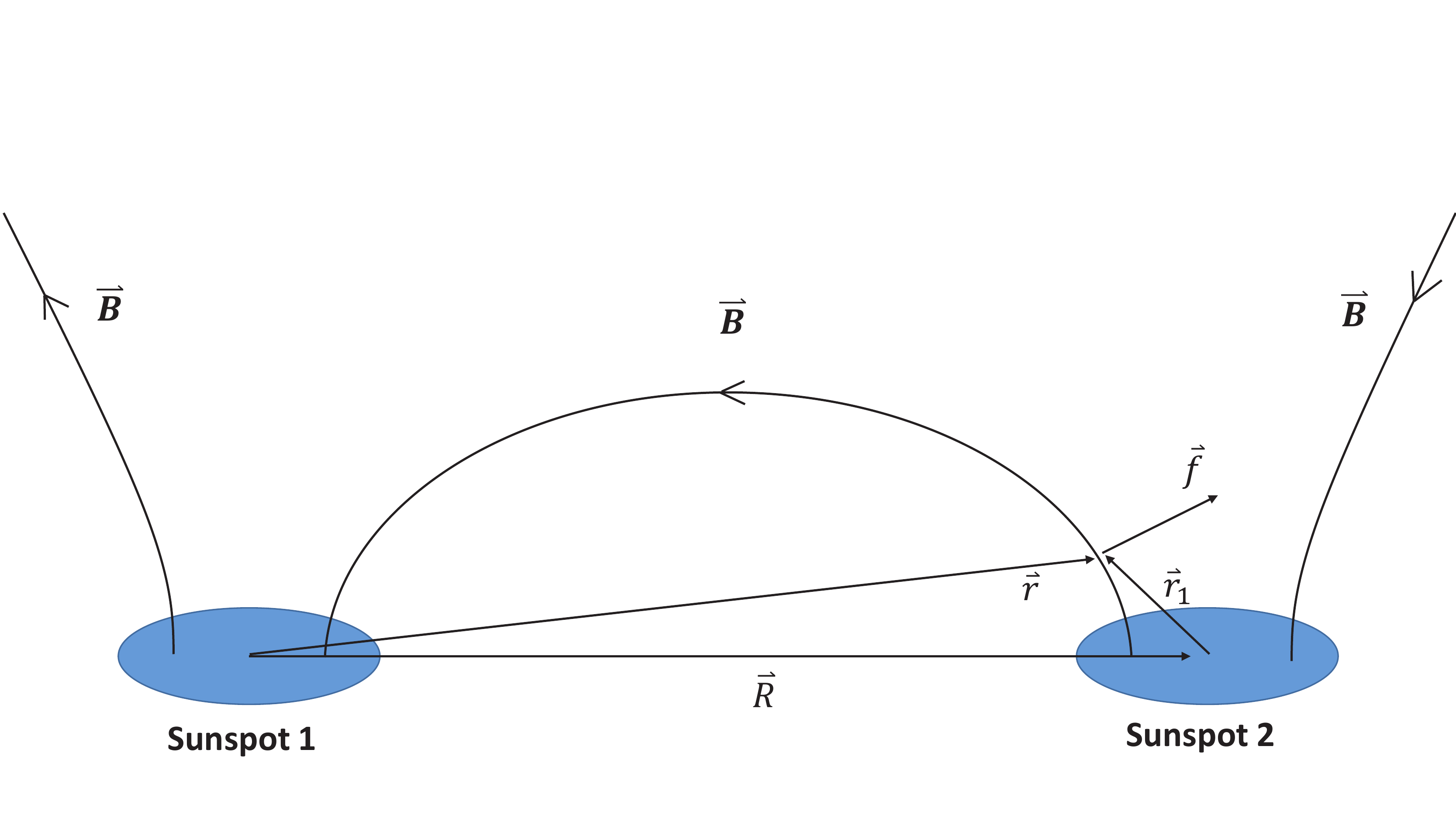}
\caption{Cartoon shows, in a MHS state, a typical interaction of the magnetic force and plasma forces in an active region with double compact polarities.}
\label{fig:fr2}
\end{figure}

Consider the case of an axisymmetric monopole sunspot (see Fig.~\ref{fig:fr}). The field line approaches a radial direction since the gas pressure becomes small. Near the spot, however, the field line has to bend to compensate the pressure gradient. According to Sec.~\ref{sec:trans}, the boundary integration of Eq.~(\ref{eq:virial3}) is invariant under the translation on the bottom plane. For the sake
of convenience, we chose ${\bf r}=0$ at the center of the spot. As gas pressure is depleted in the spot, the negative pressure gradient usually points to the interior of the spot. To compensate this force, a reversed Lorentz force is generated transversely. Then a vertically downward component of the Lorentz force is required to make sure that the Lorentz force is perpendicular to the magnetic field. The negative net Lorentz force has been confirmed in most active regions by statistical studies \citep{mcy02,t12,lsz13,lh15}. According to Fig.~\ref{fig:fr}, far away from the spot, the contribution of $\bf f\cdot r$ vanishes as the field line becomes radial. However, near the spot, the contribution of $\bf f\cdot r$ is always negative. Thus we get, in a monopole spot case,
\begin{equation}
\int_V {\bf f\cdot r} dv<0 \label{eq:fr}.
\end{equation}
The above inequality also holds in an active region with multiple spots. See Fig.~\ref{fig:fr2} with two spots, the $\bf f\cdot r$ of the left spot is equal to that of the monopole spot case. As to the right spot, with the coordinate transformation ${\bf r} = {\bf r_1}+{\bf R}$, we have
\begin{eqnarray}
&&\int_V{\bf f\cdot r}dv\\
=&&\int_V{\bf f\cdot (r_1+R)}dv\\
=&&\int_V{\bf f\cdot r_1}dv+{\bf R\cdot}\int_V{\bf f}dv\\
=&&\int_V{\bf f\cdot r_1}dv\\
<&&0,
\end{eqnarray}
where $\bf R$ is transverse while $\int_V{\bf f}dv$ is vertical. Term ${\bf R\cdot}\int_V{\bf f}dv$ vanishes is because the horizontal vector $\bf R$ is perpendicular to the vector $\int_V{\bf f}dv$, which is vertical due to the axisymmetry assumption. It is worth noting that, in the multiple spots case, the magnetic field line deviates from the radial direction because of the attraction from other spots. This effect, however, is minor near the spot. In regions far away from spots where field lines are not radial any more, the magnitude of $\bf f\cdot r$ decreases with altitude rapidly because $f$ decreases exponentially (depending on temperature) while $r$ can only grow linearly. Thus the analysis expression above still works. Therefore, we conclude that, as in an active region with sunspots, the surface integral of Eq.~(\ref{eq:virial3}) is usually an underestimate of the magnetic energy of the active region.

\section{Preprocessing method}\label{sec:preprocess}

To see if a vector magnetogram can be served as the boundary condition for a MHS system, three parameters are proposed  (similar to those for a force-free field \citep{wis06}) as follows:
\begin{eqnarray}
\epsilon_{flux}&=&\frac{|\int B_zdxdy|}{\int |B_z|dxdy}, \\
\epsilon_{force}&=&\frac{|\int B_xB_zdxdy|+|\int B_yB_zdxdy|}{\int ({B_x}^2+{B_y}^2+{B_z}^2)dxdy}, \\
\epsilon_{torque}&=&\frac{|\int (yB_zB_x-xB_zB_y)dxdy|}{\int \sqrt{x^2+y^2}({B_x}^2+{B_y}^2+{B_z}^2)dxdy}.
\end{eqnarray}
A vector magnetogram is suitable for a MHS extrapolation at least if $\epsilon_{flux}\ll 1$, $\epsilon_{force}\ll 1$ and $\epsilon_{torque}\ll 1$.

Since the vector magnetogram sometimes do not fulfill the aforementioned criteria, a preprocessing procedure is required to modify the data within the freedom of the noise. The algorithm is based on the preprocessing method that was developed by \cite{wis06}. As the restrictions on the boundary values for a MHS equilibrium are weaker compared with those for a force-free field, we need to change the preprocessing procedure accordingly. Non-zero values of z-component of net Lorentz force and x-, y-component of net torque are allowed to exist in a MHS equilibrium. Therefore these three numbers should be retained during the preprocessing process.

To do so, we define the functional
\begin{align}
\begin{split}
L&={}\mu_1L_1+\mu_2L_2+\mu_3L_3+\mu_4L_4\label{eq:L},
\end{split}\\
\begin{split}\label{eq:L1}
%L_1&={}\left(\sum_{p}B_xB_z\right)^2+\left(\sum_{p}B_yB_z\right)^2\left(a_0-\sum_{p}({B_z}^2-{B_x}^2-{B_y}^2)\right)^2,
L_1&={}\left(\sum_{p}B_xB_z\right)^2+\left(\sum_{p}B_yB_z\right)^2\\
   &\hspace{0.3em}+\left(a_0-\sum_{p}({B_z}^2-{B_x}^2-{B_y}^2)\right)^2,
\end{split}\\
\begin{split}\label{eq:L2}
%L_2&={}\left(a_1-\sum_{p}y({B_z}^2-{B_x}^2-{B_y}^2)\right)^2+\left(a_2-\sum_{p}x({B_z}^2-{B_x}^2-{B_y}^2)\right)^2+\left(\sum_{p}(yB_xB_z-xB_yB_z)\right)^2,
L_2&={}\left(a_1-\sum_{p}y({B_z}^2-{B_x}^2-{B_y}^2)\right)^2\\
   &\hspace{0.3em}+\left(a_2-\sum_{p}x({B_z}^2-{B_x}^2-{B_y}^2)\right)^2\\
   &\hspace{0.3em}+\left(\sum_{p}(yB_xB_z-xB_yB_z)\right)^2,
\end{split}\\
\begin{split}\label{eq:L3}
%L_3&={}\sum_{p}(B_x-B_{xobs})^2+\sum_{p}(B_y-B_{yobs})^2+\sum_{p}(B_z-B_{zobs})^2,
L_3&={}\sum_{p}(B_x-B_{xobs})^2+\sum_{p}(B_y-B_{yobs})^2\\
  &\hspace{0.3em}+\sum_{p}(B_z-B_{zobs})^2,
\end{split}\\
\begin{split}\label{eq:L4}
L_4&={}\sum_{p}\left((\bigtriangleup B_x)^2+(\bigtriangleup B_y)^2+(\bigtriangleup B_z)^2\right),
\end{split}
\end{align}
where
\begin{eqnarray}
a_0&=&\sum_{p}({B_{zobs}}^2-{B_{xobs}}^2-{B_{yobs}}^2), \\
a_1&=&\sum_{p}y({B_{zobs}}^2-{B_{xobs}}^2-{B_{yobs}}^2), \\
a_2&=&\sum_{p}x({B_{zobs}}^2-{B_{xobs}}^2-{B_{yobs}}^2),
\end{eqnarray}
are the z-component of net Lorentz force and the x-, y-component of net Lorentz torque, respectively. The summation is over all $p$ grid nodes on the photosphere. The weighting factors $\mu_n$ are as yet undetermined. The terms $L_1$ and $L_2$ correspond to the net force and net torque constraints. The term $L_3$ measures the deviation between the original data and the preprocessed data. The term $L_4$ controls the smoothing. For computational reasons, sufficiently smooth data are necessary for an optimization to obtain a good solution. The new algorithm differs from that for the force-free field \citep{wis06} by introducing three quantities: $a_0$, $a_1$, and $a_2$, which ensure that the preprocessing does not change the corresponding integration values.

%Normalization has been performed to the magnetic field strength and length scale with average magnetic field on the photosphere and the size of magnetogram, respectively.

%Note that in a NLFFF preprocessing, the smoothing term is included to make the magnetogram fit the smooth coronal magnetic field. In a MHS preprocessing, however, this term still exists even though we are more interested in the lower and small scale magnetic fields. It is

%Also note that the smoothing term that was used in the NLFFF preprocessing aimed to make the magnetogram fit the smooth coronal magnetic field does not appear in the functional~(\ref{eq:L1}). That is because, for the MHS extrapolation, we are more interested in the lower and small scale magnetic fields.

The strategy of preprocessing is to use the gradient descent method to minimize $L$, and meanwhile make all $L_n$ small as well. The magnetic field is optimized as follows:
\begin{equation}
minimize \quad L(B_x,B_y,B_z)
\end{equation}
with a gradient descent method
\begin{eqnarray}
{B_x}^{n+1}={B_x}^n-\delta L_{B_x},\\
{B_y}^{n+1}={B_y}^n-\delta L_{B_y},\\
{B_z}^{n+1}={B_z}^n-\delta L_{B_z}.
\end{eqnarray}
The three functional derivatives at node (q) are defined as
\begin{align}
\begin{split}
\delta L_{(B_x)_q}&={}2\mu_1\left[\left(\sum_{p} B_xB_z\right)(B_z)_q-2\left(\sum_{p}({B_z}^2-{B_x}^2-{B_y}^2)-a_0\right)(B_x)_q\right] \\
                 &\hspace{0.1em}+2\mu_2\left[2\left(a_1-\sum_{p}y({B_z}^2-{B_x}^2-{B_y}^2)\right)y(B_x)_q \right.\\
                 &\hspace{0.1em}\qquad+2\left(a_2-\sum_{p}x({B_z}^2-{B_x}^2-{B_y}^2)\right)x(B_x)_q \\
                 &\left.\hspace{0.1em}\qquad+\left(\sum_{p}(yB_xB_z-xB_yB_z)\right)y(B_z)_q\right]\\
                 &\hspace{0.1em}+2\mu_3(B_x-B_{xobs})_q2+\mu_4(\bigtriangleup(\bigtriangleup B_x))_q,
\end{split}\\
\begin{split}
\delta L_{(B_y)_q}&={}2\mu_1\left[\left(\sum_{p} B_yB_z\right)(B_z)_q-2\left(\sum_{p}({B_z}^2-{B_x}^2-{B_y}^2)-a_0\right)(B_y)_q\right] \\
                 &\hspace{0.1em}+2\mu_2\left[2\left(a_1-\sum_{p}y({B_z}^2-{B_x}^2-{B_y}^2)\right)y(B_y)_q \right.\\
                 &\hspace{0.1em}\qquad+2\left(a_2-\sum_{p}x({B_z}^2-{B_x}^2-{B_y}^2)\right)x(B_y)_q \\
                 &\left.\hspace{0.1em}\qquad-\left(\sum_{p}(yB_xB_z-xB_yB_z)\right)x(B_z)_q\right]\\
                 &\hspace{0.1em}+2\mu_3(B_y-B_{yobs})_q+2\mu_4(\bigtriangleup(\bigtriangleup B_y))_q,
\end{split}\\
\begin{split}
\delta L_{(B_z)_q}&={}2\mu_3(B_z-B_{zobs})_q+2\mu_4(\bigtriangleup(\bigtriangleup B_z))_q.
\end{split}
\end{align}
Smoothing was performed for all three components. Effects from terms that have mixed products of vertical and transverse magnetic field components in functional Eqs.~(\ref{eq:L1}-\ref{eq:L4}) are not included when evaluating $B_z$. This is designed for the fact that $B_z$ is measured with much higher accuracy than $B_x$ and $B_y$ \citep{mda11}.

%Only the transverse fields are optimized while the vertical fields are unchanged during the preprocessing. This is designed for the fact that $B_z$ is measured with much higher accuracy than $B_x$ and $B_y$.

\section{Application to \textsc{Sunrise}/IMaX data}\label{sec:application}

\begin{figure*}
%\sidecaption
\centering
\includegraphics[width=17cm]{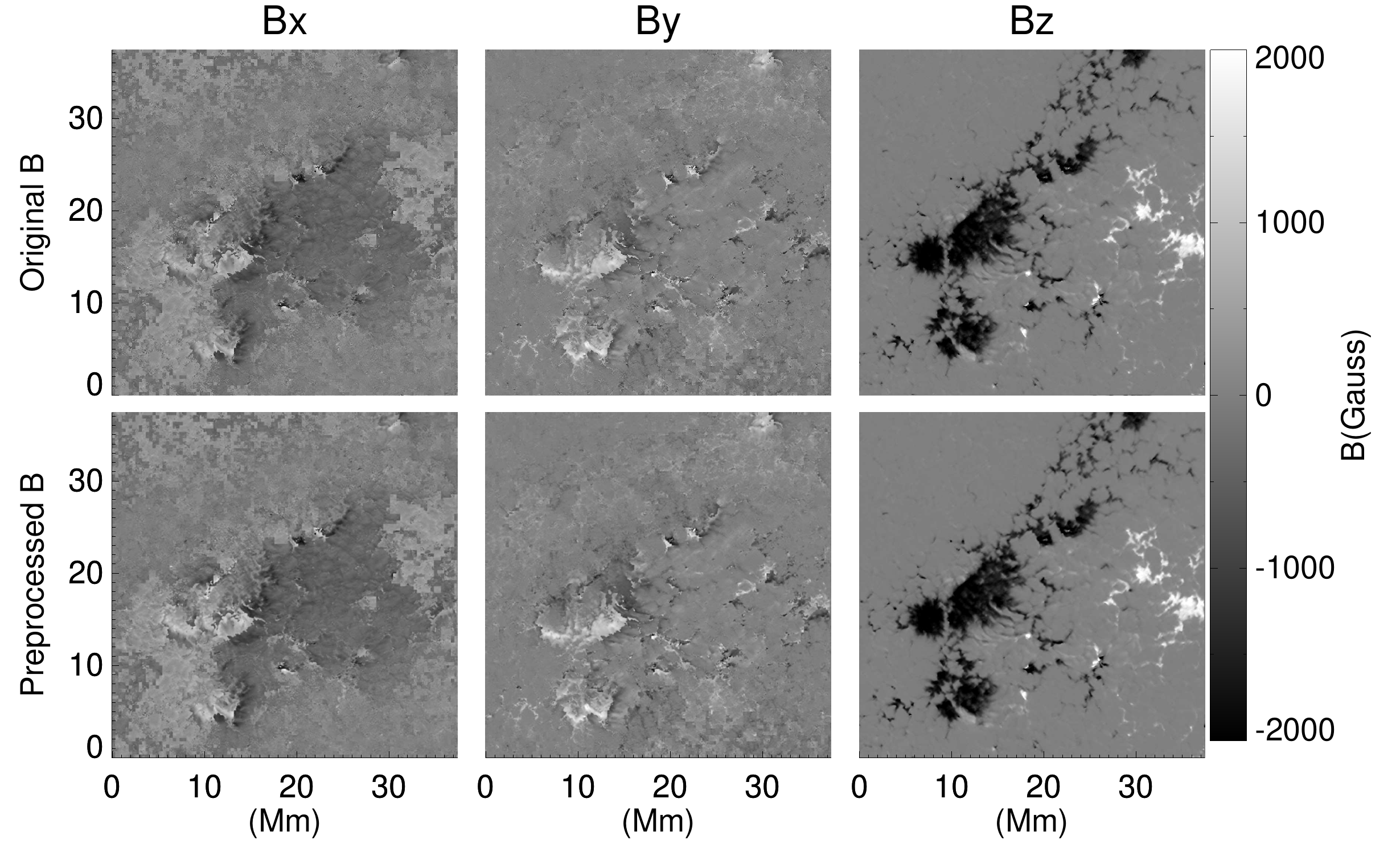}
\caption{Vector magnetogram of IMaX measured on 2013 June 12 at 23:48 UT. Top: Original data. Bottom: Preprocessed data.}
\label{fig:magnetogram}
\end{figure*}

For the test we used a combined vector magnetogram in which the \textsc{Sunrise}/IMaX data \citep{mda11,srb17} is embedded in the HMI data \citep{ssb12}. We have used this dataset to extrapolate the magnetic field as well as the plasma using two approaches \citep{wnn17,zws20}. The top panels of Fig.~\ref{fig:magnetogram} show the original vector magnetogram within the IMaX field of view (FOV). The combined data have a flux imbalance of $\epsilon_{flux}=0.013$ (almost balanced). The MHS criteria are not fulfilled to some extent with $\epsilon_{force}=0.091$ and $\epsilon_{torque}=0.066$. For comparison, Aly's criteria for force-free field are largely violated with $\epsilon_{force}=0.29$ and $\epsilon_{torque}=0.32$.

Before the magnetogram is preprocessed, we first need to choose the appropriate $\mu_n$. There are four parameters of $\mu_n$. Only three of them are independent since only the ratio of the parameters really counts. We further assume $\mu_1=\mu_2D^2\equiv\mu_{12}$, where $D$ is the average edge length of the magnetogram. Thus we could define $L_{12}=L_1+D^2L_2$. This assumption gives the same weight to the momentum and torque constraints. As only the ratio of the parameters counts, we specify $\mu_{12}=1/B_{ave}$, where $B_{ave}$ is the average magnetic field strength in the magnetogram. Then only two independent parameters remain ($\mu_3$ and $\mu_4$). A survey of the two parameters for the combined magnetogram shows that, as was also found in \cite{wis06}, $log(L_3)$ and $log(L_4)$ are almost determined by the ratio of $\mu_3$ to $\mu_4$, while $log(L_{12})$ depends on the magnitude of $\mu_3$ and $\mu_4$.

\begin{figure*}
\sidecaption
\includegraphics[width=12cm]{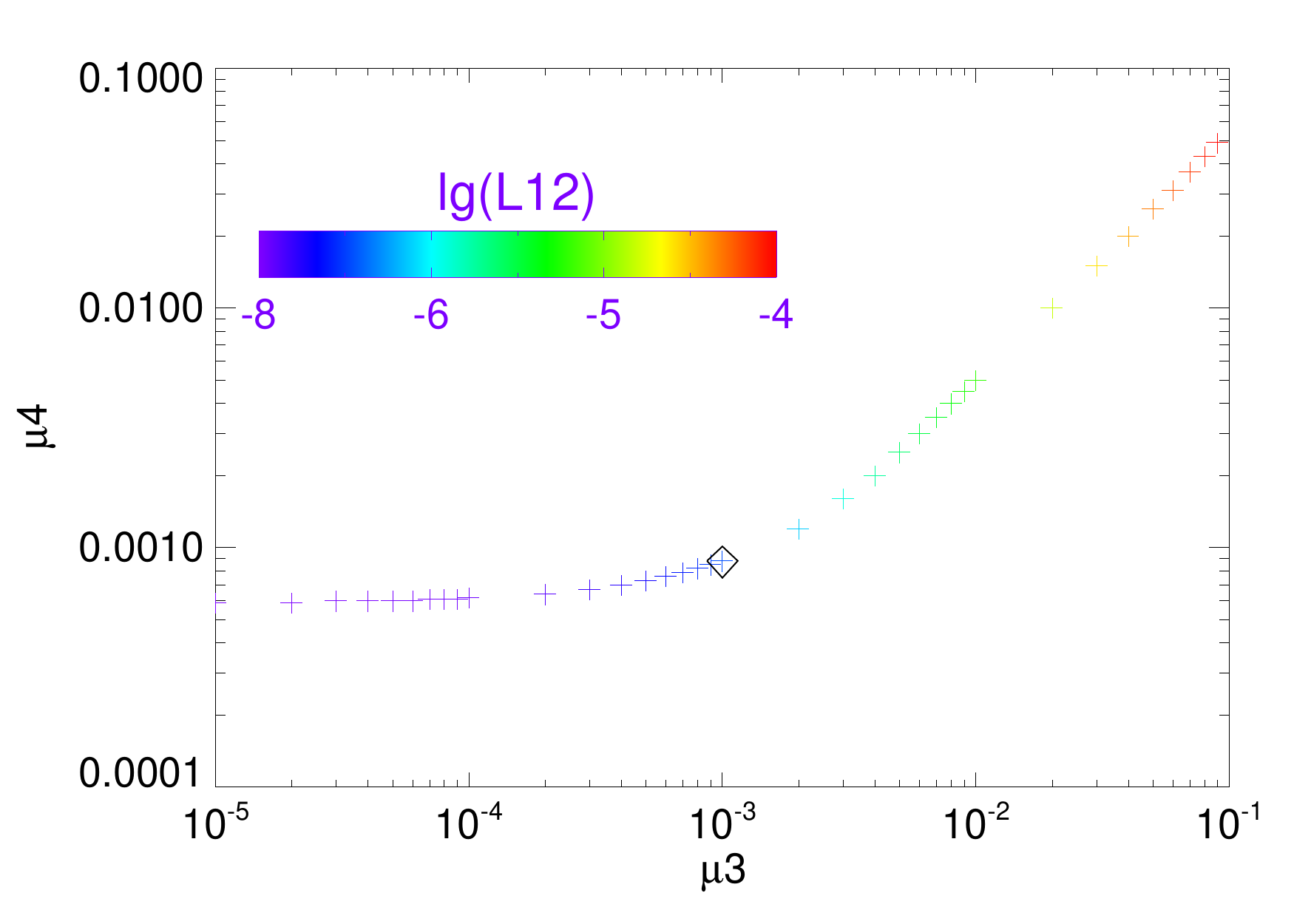}
\caption{Test runs with various $\mu_3$ and $\mu_4$ combinations. The quantity $L_3$ of each test reaches the noise level of the magnetogram. The optimal combination is denoted with a rhombus, where $\mu_3=1.0\times10^{-3}$ and $\mu_4=8.9\times10^{-4}$.}
\label{fig:mu_L}
\end{figure*}

With a deviation of the magnetic field value (i.e., a finite $L_3$), we obtain a smoothed solution that satisfies the criteria of a MHS system. The deviation is tolerable as long as $L_3$ does not exceed noise level of the magnetogram, that is, $L_3=2.2\times10^{-9}$ in this case. The noise is retrieved from the HMI dataset \citep{hlh14}. Fig.~\ref{fig:mu_L} shows optimal $\mu_3$ and $\mu_4$ combinations at which $log(L_3)$ equals to noise level of the magnetogram. As $\mu_3$ and $\mu_4$ decrease, $L_{12}$ converges. Tab.~\ref{tab:L} shows L-values as well as three summations of the initial data and the preprocessed data. The three summations are not changed during the preprocessing owing to the special design of the algorithm for fixing them. An improvement on $L_{12}$ is obvious in that $L_{12}$ is reduced by five orders of magnitude, which enforce a good compliance with the MHS criteria.\ The quantity $L_3$ has to be finite because we allow the field values to deviate from the observed values. Even though the smoothness is hardly visible in Fig.~\ref{fig:magnetogram}, $L_4$ after preprocessing is 1/20 smaller than it was before preprocessing. Now we get $\epsilon_{force}=7.0\times 10^{-5}$ and $\epsilon_{torque}=8.7\times 10^{-5}$ with an optimal values of $\mu_3=1.0\times10^{-3}$ and $\mu_4=8.9\times10^{-4}$, which are more suitable for a MHS extrapolation.

%Now we get $\epsilon_{force}=5.2\times 10^{-4}$ and $\epsilon_{torque}=5.8\times 10^{-4}$, which are more suitable for a MHS extrapolation.

%Bottom panels of Fig.~\ref{fig:magnetogram} show the magnetogram after preprocessing. The preprocessing introduces some smoothness, but retains the structures in all three components.

%We choose both small numbers ($\mu_3=7.0\times10^{-3}$ and $\mu_4=3.5\times10^{-3}$) to reduce $L_{12}$ by 4 orders of magnitude, which enforce a good compliance with the MHS criteria.

%The Linear relationship between $\mu_3$ and $\mu_4$ as well as

%These combinations are plotted in Fig.~\ref{fig:mu_L}.

%there are many combinations of $\mu_3$ and $\mu_4$ that lead to $L_3$ equals to the noise level.

%, as $\mu_3$ decreases, $L_{12}$ and $L_3$ converges to $\sim5\times10^{-8}$ and $\sim5\times10^{-10}$, respectively. We note that $L_3$ is always below the noise level of the magnetogram, which means a moderate change is enough to ensure a significant reduction of $L_{12}$. The noise information is retrieved from the HMI dataset \citep{hlh14}.

%The preprocessing introduces some smoothing, but retains the structures in all three components.

%The left panel of Fig.~\ref{fig:line} shows the magnetic field lines of a potential field reconstruction and the right panel shows those of the MHS reconstruction with the preprocessed data.

\begin{figure*}
\centering
\includegraphics[width=17cm]{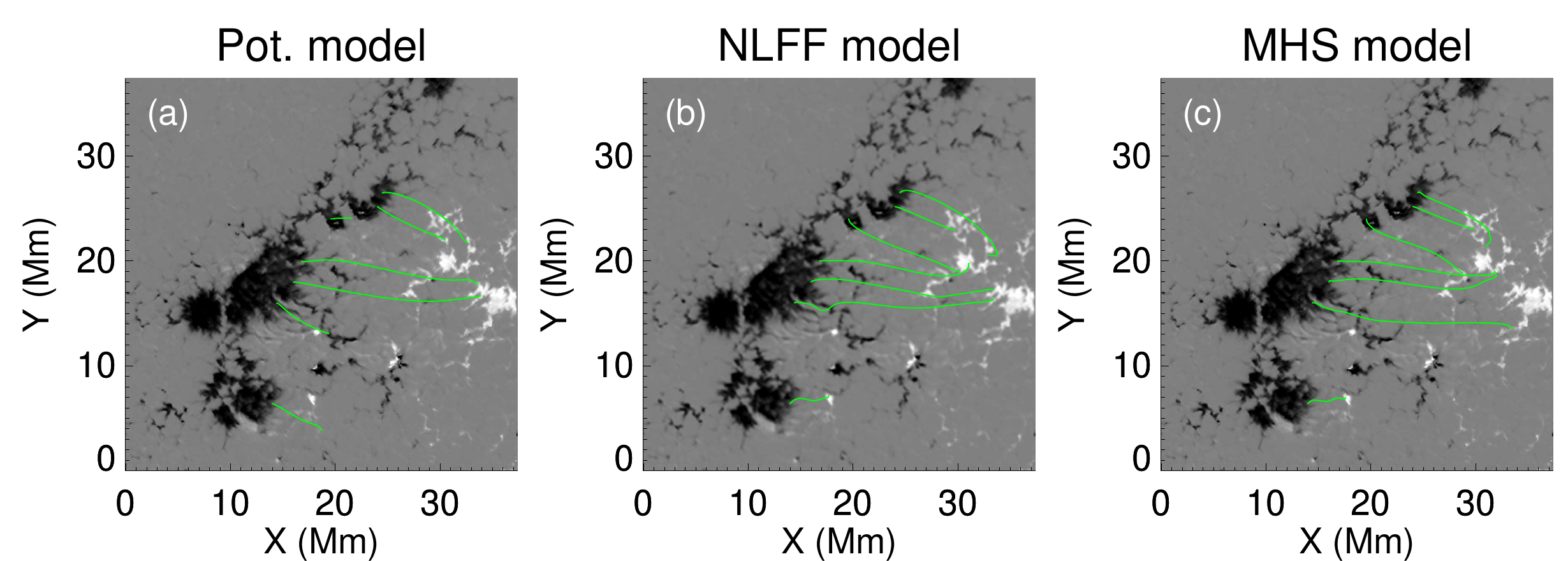}
\caption{(a) Potential field reconstruction. (b) NLFF reconstruction. (c) MHS reconstruction. The field lines originate from the same footpoints in the negative polarity of each panels.}
\label{fig:line}
\end{figure*}

Fig.~\ref{fig:line} shows field lines of different models. The original line-of-sight magnetogram was used for the potential field modeling, the force-free preprocessing was applied for the NLFFF modeling, and the MHS preprocessing was applied for the MHS modeling. All extrapolations were done in a $2336\times1824\times128$ box. A central box with $936\times936\times128$ dimensions above the IMaX FOV was cut to display the result. It is clear from fig.~\ref{fig:line} that field lines that share the same footpoints have different structures. A relatively large difference (e.g., the connectivity) can be found between panel (a) and (b) due to the force-free currents in the NLFFF. The difference between panel (b) and (c) as a consequence of the perpendicular component of the current are much smaller. Field lines in  both panel (b) and (c) have quite a similar pattern, but still have a different connectivity as well as length of field lines.

\begin{figure}
\resizebox{\hsize}{!}{\includegraphics{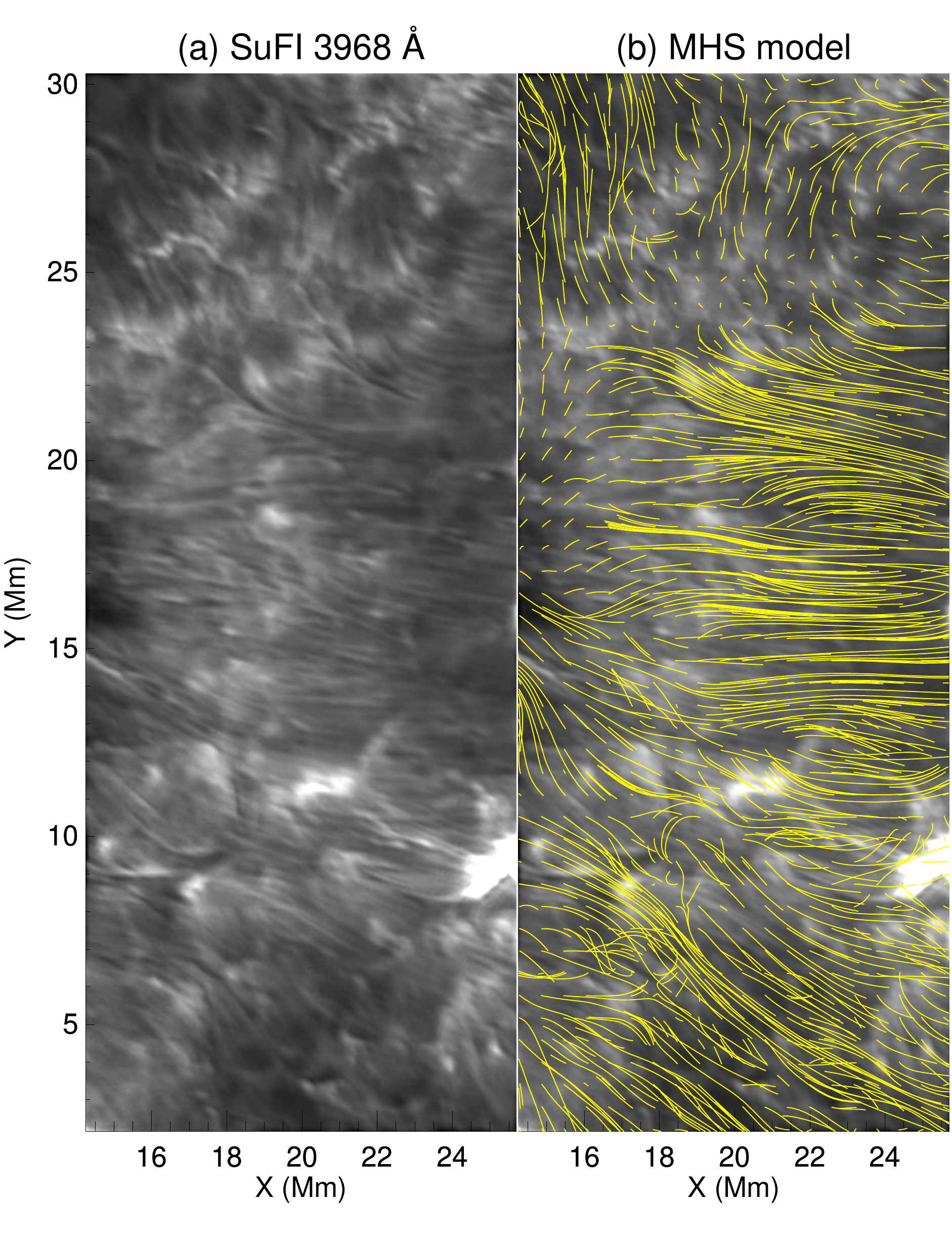}}
\caption{MHS field lines within the heights [600, 1400] km overplotted on the image observed in Ca II H core line with 1.1 $\AA$ wide filter. Footpoints of the field lines are uniformly selected on the photosphere.}
\label{fig:fl_sufi397}
\end{figure}

In Fig.~\ref{fig:fl_sufi397} we compare the simultaneous chromospheric observation by \textsc{Sunrise}/SuFI 3968 $\AA$ \citep{ggb11,srb17} with field lines within subvolumes spanning the 600-1400 km height range. We note that the extrapolation data were cut according to the SuFI FOV. The observed slender fibrils are generally believed to outline the magnetic fields in this layer. We find that most field lines trace the fibrils nicely. However, deviations can also be observed in some regions.

\section{Conclusions}\label{sec:conclusion}

We have investigated the virial theorem for a MHS system. By applying it to a large active region, we find that the surface integral that was often used to compute the energy of the force-free magnetic field is a lower bound of the magnetic energy in the semi-infinite volume. We also find a set of surface integrals as necessary criteria for a MHS system. These integrals are equivalent to Aly's criteria for a force-free field. According to the new set of criteria, we proposed an optimization algorithm to preprocess the vector magnetogram with the aim to use the result as a suitable input for a MHS extrapolation. The optimization strategy is similar to that designed in \cite{wis06} for the force-free modeling, which is to force the data to fulfill the criteria for the MHS system and be sufficiently smooth within the freedom of the measurement noise. We also show an application of the preprocessing method to a combined vector magnetogram in which the IMaX data are embedded in the HMI data.

%                                                One column figure
%-----------------------------------------------------------------

\begin{table*}
\caption{L-values and other indexes before and after preprocessing.}             % title of Table
\label{tab:L}      % is used to refer this table in the text
\centering
\begin{tabular}{c c c c c c c}        % centered columns (4 columns)
\hline\hline                 % inserts double horizontal lines
\noalign{\smallskip}
Model & $L_{12}$ & $L_3$ & $L_4$ & $\sum({B_z}^2-{B_x}^2-{B_y}^2)$ & $\sum y({B_z}^2-{B_x}^2-{B_y}^2)$ & $\sum x({B_z}^2-{B_x}^2-{B_y}^2)$ \\    % table heading
\hline                        % inserts single horizontal line
\noalign{\smallskip}
   Original  & $8.5\times10^{-3}$ & 0.0 & $1.1\times10^{-7}$ & 0.20 & 0.11 & 0.14 \\      % inserting body of the table
   \noalign{\smallskip}
Preprocessed & $2.4\times10^{-8}$ & $2.2\times10^{-9}$ & $6.4\times10^{-9}$ & 0.20 & 0.11 & 0.14 \\
   \noalign{\smallskip}
\hline                                   %inserts single line
\end{tabular}
\end{table*}

%-----------------------------------------------------------------
%\begin{appendix}
%\section{Volume integration of plasma forces} \label{app:vol}
%See Fig.~\ref{fig:fr1}, we have the integration:
%\begin{eqnarray}
%&&\int_V({\bf f\cdot r})dv\\
%=&&\int\int dRdz\left[\int_{-\frac{1}{2\pi}}^{\frac{1}{2\pi}}(-2R^2f_xcos^2\theta)d\theta+\int_{0}^{2\pi}zf_zRd\theta\right]\\
%=&&\int\int dRdz\left[\pi R(-Rf_x+2zf_z)\right].
%\end{eqnarray}
%The function $(-Rf_x+2zf_z)$ has a little difference with the RHS of Eq.~(\ref{eq:fr}), which in caused by the circular integration in a given plane. Nevertheless, the conclusion of boundary integral $\int(xB_x+yB_y)dxdy$ underestimates the magnetic energy of a large active region still holds.
%\end{appendix}

\clearpage

\begin{acknowledgements}
We thank the referee for helpful comments and suggestions. The German contribution to \textsc{Sunrise} and its reflight was funded by the Max Planck Foundation, the Strategic Innovations Fund of the President of the Max Planck Society (MPG), DLR, and private donations by supporting members of the Max Planck Society, which is gratefully acknowledged. This work was supported by DFG-grant WI 3211/4-1.
\end{acknowledgements}

   \bibliographystyle{aa} % style aa.bst
   \bibliography{aa} % your references Yourfile.bib

\end{document}